\documentclass[aps,pra,twocolumn,showkeys,showpacs]{revtex4}

\newtheorem{teorema}{Theorem}

\usepackage{amsmath}
\usepackage{amssymb}
\usepackage{dcolumn}

\begin{document}
 
\title{Lower bounds on entanglement of formation for general Gaussian states}

\author{G. Rigolin}
\email{rigolin@ifi.unicamp.br}
\author{C. O. Escobar}
\email{escobar@ifi.unicamp.br}
\affiliation{Departamento de Raios C\'osmicos e Cronologia, Instituto de F\'{\i}sica Gleb Wataghin, Universidade Estadual de Campinas, C.P. 6165, cep 13084-971, Campinas, S\~ao Paulo, Brazil}

\begin{abstract}
We derive two lower bounds on entanglement of formation for arbitrary mixed Gaussian states by two distinct methods. To achieve the first one we use a local measurement procedure that symmetrizes a general Gaussian state and the fact that entanglement cannot increase under local operations and classical communications.  The second one is obtained via a generalization to mixed states of an interesting result already known for pure states, which says that squeezed states are those that, for a fixed amount of entanglement, maximize Einstein-Podolsky-Rosen-like correlations.  
\end{abstract}



\maketitle

\section{INTRODUCTION}
The quantification of the amount of entanglement a quantum system possesses is still an open problem in Quantum Information Theory. Restricting our attention to bipartite systems, i. e., systems composed of two subsystems,  we have one measure of entanglement, entanglement of formation (EoF) \cite{bennett}, which has a clear physical meaning. Given an entangled state $\rho$, the EoF for this state expresses the number of maximally entangled states we need to create $\rho$ \cite{wootters}. The formal definition of the EoF is:

\begin{equation}
E_{f}(\rho) = \text{inf}\sum_{j}^{}p_{j}E(\psi_{j}),  \label{EoF}
\end{equation}  
where we take the infimum over all pure-state decompositions of $\rho=\sum_{j}^{}p_{j}\left|\psi_{j}\right>\left<\psi_{j}\right|$, $\sum_{j}^{}p_{j}=1$ and $E(\psi_{j})$ is the von Neumann entropy of the pure state $\psi_{j}$.

The analytical minimization of Eq.~(\ref{EoF}) is not an easy task. Dealing with two-qubit systems, which are the simplest entangled bipartite systems, Wootters \cite{wootters2} obtained an analytical expression for the EoF and Giedke \textit{et al} \cite{werner} derived an analytical expression for the EoF for symmetric Gaussian states.

Gaussian states are very useful  in quantum-optical implementation of several quantum information protocols. (Quantum cryptography \cite{gross} is an important example.) Hence, a complete characterization of the amount of entanglement of Gaussian states is desirable. The natural next step is the search for an analytical expression for the EoF for arbitrary Gaussian states. 

In this article we give two analytical expressions that furnish lower bounds for the EoF for Gaussian states. We employ two different methods to derive such lower bounds. The first lower bound is obtained using a local measurement procedure derived by Giedke \textit{et al} \cite{giedke} which symmetrizes a general Gaussian state and the fact that entanglement cannot increase under local operations and classical communications (LOCC).  The second one is derived via a generalization to mixed states of an interesting result derived by Giedke \textit{et al} \cite{werner}, who show that squeezed states are those that, for a fixed amount of entanglement, maximize Einstein-Podolsky-Rosen-like correlations. These lower bounds are also useful to rule out several possible candidates for the analytical expression of the EoF for arbitrary Gaussian states, as we illustrate in this article.
  
\section{FIRST LOWER BOUND}

Let us begin setting the notation used in this article and some properties of Gaussian states. Consider a bipartite Gaussian system $\rho$ of two modes described by the annihilation operators $a_{j}=(X_{j}+iP_{j})/\sqrt{2}$, where $j=1,2$ and $[X_{j},P_{j'}]=i\delta_{jj'}$. This system can be alternatively described by its characteristic function \cite{giedke}:
\begin{equation}
\chi(r) = \text{tr} [\rho D(r)], \label{chac}
\end{equation}  
where $r=(x_{1},p_{1},x_{2},p_{2})^{T}$ is a column real vector and 
\begin{equation}
D(r) = e^{-i(x_{1}X_{1}+p_{1}P_{1}+x_{2}X_{2}+p_{2}P_{2})}.
\end{equation}
Eq.~(\ref{chac}) uniquely defines a state $\rho$ and for Gaussian states it can always be put in the following form:
\begin{equation}
\chi(r) = e^{-\frac{1}{4}r^{T}\gamma r-id^{T}r},
\end{equation}    
where $T$ means transposition, $\gamma$ is a $4\times 4$ matrix which is called correlation matrix (CM) and $d$ is a $4-$dimensional real vector. The first moments of a Gaussian state $\left< X_{j}\right>$ and $\left<P_{j}\right>$ can always be set to zero using local unitary operations, which implies that we can work with zero mean Gaussian states when studying entanglement properties of such systems. The matrix elements $\gamma_{ij}$ of the CM can be calculated directly from the density matrix $\rho$ by the following formula:
\begin{equation}
\gamma_{ij}=\text{tr}\left[(R_{i}R_{j}+R_{j}R_{i})\rho\right]-2\text{tr}[R_{i}\rho]tr[R_{j}\rho],
\end{equation}
where $R=(X_{1},P_{1},X_{2},P_{2})^{T}$. A matrix $\gamma$ represents a realizable physical state iff it is strictly positive, real, symmetric and satisfies \cite{giedke}:
\begin{equation}
\gamma \geq J^{T}\gamma^{-1}J,
\end{equation}
where $J=\bigoplus_{k=1}^{2}J_{1}$ is a $4\times 4$ matrix with 
$J_{1}=
\left( 
\begin{array}{cc}
0 & -1 \\	
1 & 0
\end{array}
\right)$.

A Gaussian system can also be represented by its Wigner distribution $W(r)$. Assuming that we are working with a zero mean Gaussian state we have \cite{tesegiedke}:
\begin{equation}
W(r)=\frac{1}{\pi^{2}}\frac{1}{\sqrt{\text{det}\gamma_{W}}}e^{-r^{T}\gamma_{W}r}.
\end{equation} 
The CM's $\gamma$ and $\gamma_{W}$ are related by the following relation:
\begin{equation}
\gamma_{W}=J^{T}\gamma^{-1}J.\label{relation}
\end{equation}

These two CM's can be brought to the following standard form by suitable local symplectic transformations \cite{giedke}:
\begin{equation}
\gamma = 
\left( 
\begin{array}{cc}
A & C \\
C^{T} & B
\end{array}
\right), \label{gama}
\end{equation} 
where
\begin{equation}
 \begin{array}{ccc}
   A = \left(
	\begin{array}{cc}
	n & 0 \\
	0 & n
	\end{array}
	\right), & 
   B =  \left(
	\begin{array}{cc}
	m & 0 \\
	0 & m
	\end{array}
	\right), &  
   C =  \left(
	\begin{array}{cc}
	k_{x} & 0 \\
	0 & k_{p}
	\end{array}
	\right). 
 \end{array}
\end{equation}
The same set of equations apply to $\gamma_{W}$:
\begin{equation}
\gamma_{W} = 
\left( 
\begin{array}{cc}
A_{W} & C_{W} \\
C^{T}_{W} & B_{W}
\end{array}
\right),
\end{equation} 
where
\begin{equation}
 \begin{array}{cc}
   A_{W} = \left(
	\begin{array}{cc}
	N & 0 \\
	0 & N
	\end{array}
	\right), & 
   B_{W} =  \left(
	\begin{array}{cc}
	M & 0 \\
	0 & M
	\end{array}
	\right),
 \end{array} 
\end{equation}
\begin{equation}  
   C_{W} =  \left(
	\begin{array}{cc}
	K_{x} & 0 \\
	0 & K_{p}
	\end{array}
	\right). 
\end{equation}
The four real parameters $(n,m,k_{x},k_{p})$ completely characterize a two mode Gaussian state and they are related to the four local symplectic transformation invariants as follows \cite{simon}:
\begin{subequations}
\label{i}
\begin{eqnarray} 
I_{1} = n = \sqrt{\text{det}A}, & \, \\ 
I_{2} = m = \sqrt{\text{det}B}, & \, \\ \label{i1}
I_{3} = k_{x}k_{p} = \text{det}C, & \,\\ \label{i3}
I_{4} = nm(k_{x}^{2}+k_{p}^{2}) = & \text{tr}\left(AJ_{1}^{T}CJ_{1}^{T}BJ_{1}^{T}C^{T}J_{1} \right). \label{i4} 
\end{eqnarray}
\end{subequations} 
Alternatively the four real parameters $(N,M,K_{x},K_{p})$ also completely specify a two mode Gaussian system. They can be also obtained by local symplectic transformation invariants. These invariants, which we call $W_{1}, W_{2}, W_{3}$ and $W_{4}$, satisfy Eq.~(\ref{i}), where we change $A, B$ and $C$ by $A_{W}, B_{W}$ and $C_{W}$ and $(n,m,k_{x},k_{p})$ by  $(N,M,K_{x},K_{p})$.

We now pass to the derivation of the first lower bound. A symmetric Gaussian entangled state $\sigma$ is completely specified by its CM (see Eq.~(\ref{gama})), where $n=m=\tilde{n}$. (From now on, every parameter associated with a symmetric Gaussian state will be represented by a tilde on top of it.) Let us assume, without loss of generality, $\tilde{k}_{x} > 0$ and $\tilde{k}_{p} < 0$ \cite{simon}. The EoF for this symmetric state is \cite{werner}:
\begin{equation}
E_{f}(\sigma) = f\left[\sqrt{(\tilde{n}-|\tilde{k}_{x}|)(\tilde{n}-|\tilde{k}_{p}|)}\right], \label{efs}  
\end{equation}
where,
\begin{equation}
f(\delta)=c_{+}(\delta)\,\log_{2}[c_{+}(\delta)]-c_{-}(\delta)\,\log_{2}[c_{-}(\delta)].
\end{equation}
Here $c_{\pm}=(\delta^{-1/2}\pm \delta^{1/2})^2/4$.
Using Eq.~(\ref{i}) we can write the EoF given by Eq.~(\ref{efs}) in terms of invariants:
\begin{equation}
E_{f}(\sigma)=f\left[ \sqrt{\tilde{I}_{1}-\tilde{I}_{3}-\sqrt{\tilde{I}_{4}-2\tilde{I}_{1}\tilde{I}_{3}}} \right]. \label{um}        
\end{equation}
Using Eqs.~(\ref{relation}) and (\ref{i}) we obtain the following relations among the invariants of the $\gamma$ and $\gamma_{W}$ matrices:
\begin{equation}
I_{1}=\frac{W_{2}}{W_{5}}, \,  I_{2}=\frac{W_{1}}{W_{5}}, \,  I_{3}=\frac{W_{3}}{W_{5}}, \,  I_{4}=\frac{W_{4}}{W_{5}^{2}}, \, I_{5}=\frac{1}{W_{5}}, \label{relacao} 
\end{equation}
where $W_{5}=\text{det}\gamma_{W}$ and $I_{5}=\text{det}\gamma$.

Therefore, due to Eq.~(\ref{relacao}) the EoF for our symmetric Gaussian state, Eq.~(\ref{um}), can be expressed as:
\begin{equation}
E_{f}(\sigma)=f\left[ \sqrt{\frac{\tilde{W}_{1}-\tilde{W}_{3}-\sqrt{\tilde{W}_{4}-2\tilde{W}_{1}\tilde{W}_{3}}}{\tilde{W}_{5}}} \right]. \label{dois}      
\end{equation}
But Giedke \textit{et al} \cite{giedke} have shown that a general bipartite Gaussian system $\rho$ can be transformed to a symmetrical bipartite Gaussian system $\sigma$ using LOCC. This implies that $E_{f}(\rho)\geq E_{f}(\sigma)$. Schematically we have: 
\begin{equation}
\rho \stackrel{LOCC}{\longrightarrow} \sigma \Longrightarrow E_{f}(\rho) \geq E_{f}(\sigma). \label{locc}
\end{equation}
Our only task now is to rewrite Eq.~(\ref{dois}) in terms of the invariants of the $\gamma$ matrix of $\rho$. It is in this step that we use Giedke's symmetrization procedure. 

Given a general bipartite Gaussian system $\rho$ and its $\gamma_{W}$ matrix, where we assume, without loss of generality that $N > M$, we can achieve by means of local operations a symmetric state with the following $\tilde{\gamma}_{W}$ matrix \cite{giedke}:
\begin{equation}
\gamma_{W} = 
\left( 
\begin{array}{cc}
A_{W} & C_{W} \\
C^{T}_{W} & B_{W}
\end{array}
\right),
\stackrel{LOCC}{\longrightarrow}
\tilde{\gamma}_{W} = 
\left( 
\begin{array}{cc}
\tilde{A}_{W} & \tilde{C}_{W} \\
\tilde{C}^{T}_{W} & \tilde{B}_{W}
\end{array}
\right),
\end{equation}
where
\begin{equation}
\tilde{A}_{W}  =  \left(
\begin{array}{cc}
\frac{N\cos^{2}\theta+(NM-K_{x}^{2})\sin^{2}\theta}{\cos^{2}\theta+M\sin^{2}\theta} & 0 \\
0 & \frac{N\cos^{2}\theta+NM\sin^{2}\theta}{\cos^{2}\theta+M\sin^{2}\theta}
\end{array}
\right), \label{de}
\end{equation}
\begin{equation}
\tilde{B}_{W}  =  \left(
\begin{array}{cc}
\frac{M}{\cos^{2}\theta+M\sin^{2}\theta} & 0 \\
0 & \sin^{2}\theta+M\cos^{2}\theta
\end{array}
\right),
\end{equation}
\begin{equation}
\tilde{C}_{W}=\left(
\begin{array}{cc}
\frac{K_{x}\cos\theta}{\cos^{2}\theta+M\sin^{2}\theta} & 0 \\
0 & K_{p}\cos\theta
\end{array}
\right), \label{ate}
\end{equation}
\begin{equation}
\tan^{2}\theta=\frac{N^{2}-M^{2}}{M-N(NM-K_{x}^{2})}. \label{condicao}
\end{equation}
Eq.~(\ref{condicao}) guarantees that det$\tilde{A}_{W}$ = det$\tilde{B}_{W}$. This condition is the statement that the Gaussian system with the $\tilde{\gamma}_{W}$ above is symmetrical \cite{giedke}.

Using Eqs.~(\ref{i},\ref{de}-\ref{ate}) and the assumption that $|K_{x}|\geq |K_{p}|$ \cite{nota1} we can write Eq.~(\ref{dois}) as follows:
\begin{equation}
E_{f}(\sigma)=f\left[ \sqrt{\frac{\alpha -\sqrt{\beta}}{\epsilon}} \right], \label{tres}      
\end{equation}
where
\begin{equation}
\alpha = W_{2}-W_{3}+\sqrt{W_{2}}\tan^{2}\theta, \label{alfa}
\end{equation}
\begin{equation}
\epsilon = W_{5} + \sqrt{W_{1}}(\sqrt{W_{1}W_{2}}-K_{x}^{2})\tan^{2}\theta,
\end{equation}
\begin{eqnarray}
\beta & = & W_{4}-2W_{2}W_{3}+ \tan^{2}\theta \left[ (W_{4}-2W_{3}-W_{3}^{2})\sqrt{W_{2}}\right.\nonumber \\ 
 & & \left. +(1-W_{2})K_{x}^{2}\sqrt{W_{1}}\right],
\end{eqnarray}
\begin{equation}
\tan^{2}\theta = \frac{W_{1}-W_{2}}{\sqrt{W_{2}}-\sqrt{W_{1}}(\sqrt{W_{1}W_{2}}-K_{x}^{2})}, 
\end{equation}
\begin{equation}
K_{x}^{2}=\frac{W_{4}+\sqrt{W_{4}^{2}-4W_{1}W_{2}W_{3}^{2}}}{2\sqrt{W_{1}W_{2}}}.
\end{equation}
Now Using Eq.~(\ref{relacao}) we can put Eq.~(\ref{tres}) in terms of the invariants of the $\gamma$ matrix. Hence, if we work with $\gamma$ in its standard form given by Eq.~(\ref{gama}), where we assume, without loss of generality, that $|k_{x}|\geq |k_{p}|$, Eq.~(\ref{tres}) is rewritten after a tedious but straightforward algebraic manipulation as \cite{nota2}:
\begin{widetext}
\begin{equation}
E_{f}(\sigma) = f\left[\sqrt{\frac{nmh(n,m)-k_{x}k_{p}h(m,n)+|mk_{x}-nk_{p}|\sqrt{h(n,m)h(m,n)}}{g(n,m)}}\right], \label{quase}
\end{equation}
\end{widetext}
where
\begin{subequations}
\label{hg}
\begin{eqnarray}
h(n,m) & = & n-m(nm-k_{p}^{2}) \\
g(n,m) & = & m(1-m^{2})+nk_{p}^{2}.
\end{eqnarray}
\end{subequations}
Eq.~(\ref{quase}) is our first lower bound for the EoF for general Gaussian states. It is worthy noting that this lower bound reduces to the EoF for symmetric Gaussian states whenever $n=m$.

\section{SECOND LOWER BOUND}

A two-mode squeezed state \cite{werner} is a symmetric Gaussian pure state that belongs to the Hilbert space $\mathcal{H} = \mathcal{H}_{1} \otimes \mathcal{H}_{2}$ and is described by the following vector:
\begin{equation}
\left|\Psi_{s}(r)\right>=\frac{1}{\cosh (r)}\sum_{n=0}^{\infty}\tanh^{n}(r) \left|n\right>_{1} \otimes \left|n\right>_{2}, \label{squeezedpure} 
\end{equation}
where $\left|n\right>_{j}$ is the n-th Fock state, that is, $a^{\dagger}_{j}a_{j}\left|n\right>_{j}=n\left|n\right>_{j}$, j=1,2 and $r \in (0,\infty)$ is the squeezing parameter.  

There exists an interesting relation between squeezed states and EPR-correlations, which Giedke \text{et al} \cite{werner} proved in their proposition $1$: Given a squeezed state $\left|\Psi_{s}(r)\right>$  and an arbitrary pure two-mode state $\left|\psi\right>$ then, if they have the same EPR-correlation, the squeezed state is the least entangled. In other words, if we call $\Delta(\Psi_{s}(r))$ and $\Delta(\psi)$ the EPR-correlations for the two mentioned states and if $\Delta(\Psi_{s}(r))=\Delta(\psi)$ then, $E(\psi)\geq E(\Psi_{s}(r))$.

The EPR-correlation is defined as \cite{werner}:
\begin{equation}
\Delta(\psi)=\text{min}\left\{ 1, \frac{1}{2}\left[ \Delta^{2}_{\psi}(X_{1}-X_{2})+\Delta_{\psi}^{2}(P_{1}+P_{2}) \right]  \right\}, \label{derta}
\end{equation}
where $\Delta^{2}_{\psi}(R_{j})=\left<R_{j}^{2}\right>_{\psi}-\left<R_{j}\right>_{\psi}^{2}$ is the dispersion of the observable $R_{j}$. The above expression measures the degree of non-local correlations, and is zero for the original EPR-state \cite{werner,epr}. This means that the more a system is non-local the more Eq.~(\ref{derta}) approaches zero. We say that a system with the minimal $\Delta(\psi)$ has the maximal EPR-correlation. For our squeezed state the EPR-correlation is \cite{werner}:
\begin{equation}
\Delta[\Psi_{s}(r)]=e^{-2r}. \label{eprsqueezed}
\end{equation}

The EoF, which is equal to the von Neumann entropy, for the squeezed state is \cite{werner}: 
\begin{eqnarray}
E[\Psi_{s}(r)] & = & \cosh^{2}(r)\log_{2}[\cosh^{2}(r)] \nonumber \\
& & -\sinh^{2}(r)\log_{2}[\sinh^{2}(r)].
\end{eqnarray}
And it is shown that \cite{werner}:
\begin{equation}
E[\Psi_{s}(r)]=f(\Delta[\Psi_{s}(r)]). \label{essa1}
\end{equation}
Giedke \textit{et al} \cite{werner} have shown that $f:(0,1] \rightarrow [0,\infty)$ is a convex and decreasing function of its argument. Hence, as Eq.~(\ref{eprsqueezed}) can have any value between zero and one, the EoF for a squeezed state can assume any value between zero and infinity. This property of the EoF for squeezed states, i. e., that they can assume any value, is an essential ingredient in our generalization of Giedke's \textit{et al} \cite{werner} proposition 1. Let us now state and then prove the following theorem which is a generalization to mixed states of Giedke's \textit{et al} \cite{werner} proposition $1$.
\begin{teorema}
For all bipartite Gaussian systems $\rho$ we have $E_{f}(\rho)$ $\geq$ $E_{f}(\sigma)$, if $\Delta(\rho) = \Delta(\sigma)$  and $\sigma$ is a symmetric Gaussian mixed state. 
\end{teorema}
Here $\Delta(\rho)$ is analogously defined as in Eq.~(\ref{derta}). \\
\textit{Proof:} Applying a suitable symplectic local transformation in the standard form of the $\gamma$ matrix of $\sigma$ \cite{werner, nota4} we see that the EPR-correlation for this transformed matrix is $\Delta(\sigma)=\sqrt{\left(\tilde{n}-|\tilde{k}_{x}|\right)\left(\tilde{n}-|\tilde{k}_{p}|\right)}$. But the amount of entanglement is invariant by local symplectic transformations. This means that  $E_{f}(\sigma)=f\left[\Delta(\sigma)\right]=f\left[\Delta(\rho)\right]$. Let us write $\rho$ as
\begin{equation}
\rho = \sum_{j}^{}p_{j}\left|\varphi_{j}\right>\left<\varphi_{j}\right|,
\end{equation}  
where the above decomposition is the one that furnishes the EoF of $\rho$, i. e.,
\begin{equation}
E_{f}(\rho)=\sum_{j}p_{j}E(\varphi_{j}).
\end{equation}
Using the above expansion of $\rho$ we have that 
\begin{eqnarray}
E_{f}(\sigma) & = & f\left[\Delta\left(\sum_{j}^{}p_{j}\left|\varphi_{j}\right>\left<\varphi_{j}\right|\right)\right] \nonumber \\ 
& & \leq f\left[\sum_{j}^{}p_{j}\Delta(\varphi_{j}) \right] \nonumber \\
& & \leq \sum_{j}^{}p_{j}f\left[ \Delta(\varphi_{j}) \right]. \label{vamola}
\end{eqnarray}
The first inequality is a consequence of the concavity of $\Delta(\rho)$ (see Appendix) and the fact that  $f$ is a decreasing function of its argument \cite{werner}. The second inequality is due to the convexity of $f$ \cite{werner}. We now use the fact that a squeezed state can assume any value of entanglement. For each pure state in the decomposition of $\rho$ above we associate a squeezed state with the same amount of entanglement: $E(\varphi_{j})=E[\Psi_{s}(r_{j})]$. Therefore we have the following relation for the EoF of $\rho$:
\begin{eqnarray}
E_{f}(\rho) & = & \sum_{j}p_{j}E(\varphi_{j}) = \sum_{j}p_{j}E[\Psi_{s}(r_{j})] \nonumber \\
& = & \sum_{j}p_{j}f\left[\Delta[\Psi_{s}(r_{j})]\right]. \label{importante}
\end{eqnarray} 
Now due to the proposition $1$ of Giedke \textit{et al} \cite{werner} we know that $\Delta(\varphi_{j}) \geq \Delta[\Psi_{s}(r_{j})]$. Hence, using this fact in Eq.~(\ref{vamola}) and that $f$ is a decreasing function of its argument we have:
\begin{equation}
E_{f}(\sigma)\leq \sum_{j}^{}p_{j}f\left[ \Delta(\varphi_{j}) \right] \leq \sum_{j}^{}p_{j}f\left[ \Delta\left[\Psi_{s}(r_{j})\right] \right]. \label{acabei}
\end{equation} 
Combining Eqs.~(\ref{importante}) and (\ref{acabei}) we see that
\begin{equation}
E_{f}(\sigma) \leq E_{f}(\rho). \label{tiurema}\; \square  
\end{equation}
The above theorem tells us that for mixed states the symmetric states are those with less EoF given an EPR-correlation. It is interesting to note that $\sigma$ can be any symmetric state, including symmetric states written as superpositions of squeezed states. 

The previous theorem automatically gives us a lower bound for the EoF for general Gaussian states. Using Eq.~(\ref{tiurema}) we get:
\begin{equation}
E_{f}(\rho) \geq E_{f}(\sigma) = f\left[ \Delta(\rho) \right]. \label{tamoquase} 
\end{equation}
We now implement a local symplectic transformation in the $\gamma$ matrix of $\rho$, Eq.~(\ref{gama}), before calculating the EPR-correlation. (It does not alter the amount of entanglement, since it is equivalent to a unitary local transformation in the density matrix $\rho$.) This transformation can be viewed as an extension to non-symmetrical Gaussian states of the transformation introduced by Giedke \textit{et al} \cite{werner} for symmetric states. This transformation multiplies $X_{j}$ by $[(n+m)/2-|k_{p}|]/[(n+m)/2-|k_{x}|]^{1/4}$. $P_{j}$ is divided by the same quantity. Now calculating $\Delta(\rho)$ we get the following expression for our second lower bound:
\begin{widetext}
\begin{equation}
E_{f}(\rho) \geq f\left[\text{min}\left\{1,\sqrt{\left(\frac{n+m}{2}-|k_{x}|\right)\left(\frac{n+m}{2}-|k_{p}|\right)}\right\}\right]. \label{lb2}
\end{equation} 
\end{widetext}
Again we see that this lower bound reduces to the EoF for symmetric systems whenever $n=m$. It is important to note that when $\sqrt{\left(\frac{n+m}{2}-|k_{x}|\right)\left(\frac{n+m}{2}-|k_{p}|\right)}>1$ we have $\Delta(\rho)=1$. For such cases this lower bound is not useful since it simply shows that $E_{f}(\rho)\geq 0$.

\section{DISCUSSION AND EXAMPLES}
We now employ the two lower bounds derived previously, Eqs.~(\ref{quase}) and (\ref{lb2}), to see their usefulness in analyzing some Gaussian states. For completeness we present in terms of the invariants $(n,m,k_{x},k_{p})$ three inequalities they must satisfy to be considered parameters that describe physically realizable entangled Gaussian states \cite{giedke}. We will assume, without loss of generality $m\geq n$ and $|k_{x}|\geq |k_{p}|$.
\begin{subequations}
\label{todas}
\begin{eqnarray}
\text{det}\gamma + 1 & \geq & n^{2} + m^{2} + 2k_{x}k_{p} \\
nm - k_{x}^{2} & \geq & 1 \\
\text{det}\gamma + 1 & < & n^{2} + m^{2} - 2k_{x}k_{p}.
\end{eqnarray} 
\end{subequations}
The last inequality is the restriction a $\gamma$ matrix must satisfy to represent an entangled Gaussian system.

The table below shows six entangled Gaussian systems and the values of their two lower bounds (LB1 and LB2).
\begin{table}[h]
\caption{\label{tabela} The first column shows the parameters of the $\gamma$ matrix when written in its standard form. The second and third column represent the two lower bounds for the EoF for mixed Gaussian states. Lower bound $1$ is given by Eq.~(\ref{quase}) and lower bound $2$ is given by Eq.~(\ref{lb2}).}
\begin{ruledtabular}
\begin{tabular}{ccc}
$n,m,k_{x},k_{p}$ & LB1 & LB2 \\
1.5, 2, 1.2, -1 & 0.14635 & 0.28919 \\
1.5, 2, 1, -1 & 0.08687 & 0.14672 \\
2, 3, 1.8, -1.2  & 0.02448 & 0.00681 \\
1.7, 2.6, 1.3, -0.9 & 0.00549 & 0 \\
2, 3, 1.7, -1.2 & 0.00725 & 0.00142 \\
2, 2.5, 1.3, -1.2 & 0.00173 & 0.00001 
\end{tabular}
\end{ruledtabular}
\end{table}
These six Gaussian systems are very representative. Looking at their lower bounds we see that depending on the parameters of the system LB1 or LB2 is the strongest lower bound. For example, the first two Gaussian systems have LB2 as the strongest lower bound but the four last Gaussian systems have LB1 as the strongest one.
LB1 and LB2 are also useful for discarding possible candidates for the EoF of a general mixed Gaussian state. Consider, just for illustration, the functions 
\begin{equation}
f_{1} =  f\left[\sqrt{\left( \sqrt{nm}-|k_{x}|\right)\left(\sqrt{nm}-|k_{p}|\right)}\right], 
\end{equation}
\begin{equation}
f_{2}  =  f\left[\sqrt{\left( \sqrt{\frac{n^{2}+m^{2}}{2}}-|k_{x}|\right)\left(\sqrt{\frac{n^{2}+m^{2}}{2}}-|k_{p}|\right)}\right]. 
\end{equation}
Both $f_{1}$ and $f_{2}$ reduce to the EoF for symmetric states when $n=m$. For the Gaussian states with $(n,m,k_{x},k_{p})$ $=$ $(2,2.5,1.3,-1.2)$ we have LB1 $=$ $0.00173$ $>$ $f_{1}$ $=$ $0.00091$ and for $(n,m,k_{x},k_{p})$ $=$ $(1.5,2,1.1,-1)$ we get LB2 $=$ $0.208853$ $>$ $f_{2}$ $=$ $0.18621$. These results show that $f_{1}$ and $f_{2}$ cannot be proved to be the EoF for general Gaussian systems since we have lower bounds for the EoF that are greater than $f_{1}$ and $f_{2}$.

\section{CONCLUSION}

We presented in this article two lower bounds for the EoF of general Gaussian two-mode systems. They were obtained by two distinct methods. 

The first lower bound, Eq.~(\ref{quase}), was derived using an interesting procedure derived by Giedke \textit{et al} \cite{giedke} that symmetrizes by local operations a Gaussian state and the well known fact that entanglement does not increase under LOCC. 

The second lower bound, Eq.~(\ref{lb2}), is a corollary of theorem $1$, which can be interpreted as an extension of a previous result obtained by Giedke \textit{et al} \cite{werner}: given two pure bipartite systems with the same amount of entanglement, the squeezed states are those with the maximal EPR-correlation. Our theorem generalizes this fact to mixed states in the sense that symmetric Gaussian states are shown to be states with maximal EPR-correlation for a fixed amount of entanglement.

\begin{acknowledgments}
The authors would like to express their gratitude to the funding of Funda\c{c}\~ao de Amparo \`a Pesquisa do Estado de S\~ao Paulo (FAPESP) and to Conselho Nacional de Desenvolvimento Cient\'{\i}fico e Tecnol\'ogico (CNPq). 
\end{acknowledgments}

\appendix*

\section{Proof of concavity of $\Delta(\rho)$}
We need to prove that $\Delta(\rho) \geq \sum_{j}^{}p_{j}\Delta(\phi_{j})$, where $\rho = \sum_{j}^{}p_{j}\left|\phi_{j}\right>\left<\phi_{j}\right|$. Applying the definition of $\Delta(\rho)$ we obtain for the given expansion of $\rho$ and for $\sum_{j}^{}p_{j}\Delta(\phi_{j})$ the following expressions:
\begin{widetext}
\begin{equation}
\Delta(\rho) =  \text{min}\left\{1,\frac{1}{2}\left[\sum_{j}^{}p_{j}\left(\left<X^{2}\right>_{\phi_{j}}+\left<P^{2}\right>_{\phi_{j}}\right)-\left(\sum_{j}^{}p_{j}\left<X\right>_{\phi_{j}}\right)^{2}-\left(\sum_{j}^{}p_{j}\left<P\right>_{\phi_{j}}\right)^{2}\right]\right\}, \label{a1}
\end{equation}
\begin{eqnarray}
\sum_{j}^{}p_{j}\Delta(\phi_{j}) & = & \sum_{j}^{}p_{j}\text{min}\left\{1,\frac{1}{2}\left[\left<X^{2}\right>_{\phi_{j}}+\left<P^{2}\right>_{\phi_{j}}-\left<X\right>^{2}_{\phi_{j}}-\left<P\right>^{2}_{\phi_{j}}\right]\right\} \label{a2} \\
 & \leq & \frac{1}{2}\sum_{j}^{}p_{j}\left[\left<X^{2}\right>_{\phi_{j}}+\left<P^{2}\right>_{\phi_{j}}-\left<X\right>^{2}_{\phi_{j}}-\left<P\right>^{2}_{\phi_{j}}\right], \label{a3}
\end{eqnarray}
\end{widetext}
where $X=X_{1}-X_{2}$ and $P=P_{1}+P_{2}$. The inequality is a consequence of the fact that we may have at least one $\left<X^{2}\right>_{\phi_{j}}+\left<P^{2}\right>_{\phi_{j}}-\left<X\right>^{2}_{\phi_{j}}-\left<P\right>^{2}_{\phi_{j}} > 2$. Looking at Eq.~(\ref{a2}) we see that it is not greater than $1$. Thus, if Eq.~(\ref{a1}) is equal to $1$ we see that $\Delta(\rho) \geq \sum_{j}^{}p_{j}\Delta(\phi_{j})$. But if it is less than $1$, $\Delta(\rho) \geq \sum_{j}^{}p_{j}\Delta(\phi_{j})$ if the following inequality is satisfied: 
\begin{widetext} 
\begin{equation} 
\left(\sum_{j}^{}p_{j}\left<X\right>_{\phi_{j}}\right)^{2}+\left(\sum_{j}^{}p_{j}\left<P\right>_{\phi_{j}}\right)^{2} \leq  \sum_{j}^{}p_{j}\left[\left<X\right>^{2}_{\phi_{j}}+\left<P\right>^{2}_{\phi_{j}}\right]. \label{a55}
\end{equation}
\end{widetext}
Applying the Cauchy-Schwarz inequality \cite{duan} for an observable $R$ we get $\sum_{j}^{}p_{j}\left<R\right>_{\phi_{j}}^{2}\geq \left( \sum_{j}^{}p_{j}\left<R\right>_{\phi_{j}}\right)^{2}$. Hence, Eq.~(\ref{a55}) is always satisfied. $\square$


\begin{thebibliography}{25}
\bibitem{nota1} The assumption that $|K_{x}|\geq |K_{p}|$ is just for convenience. We do not lose in generality by such assumption. We need it to correctly express $K_{x}^{2}$ in terms of the $\gamma_{W}$ invariants, i. e., $K_{x}^{2}=\frac{W_{4}+\sqrt{W_{4}^{2}-4W_{1}W_{2}W_{3}^{2}}}{2\sqrt{W_{1}W_{2}}}$. If we have assumed that  $|K_{x}|< |K_{p}|$ we would have gotten a minus sign before the square root.
\bibitem{nota2} We have $m\geq n$ in the standard form of $\gamma$ because we have assumed $N \geq M$ in the standard form for $\gamma_{W}$. This can be seen to be true by looking at Eq.~(\ref{relacao}), where we have $n^{2}=M^{2}/\text{det}\gamma_{W}$ and $m^{2}=N^{2}/\text{det}\gamma_{W}$.
\end{thebibliography}
\end{document}